%% file: main.tex
\documentclass[nonacm,sigconf]{acmart}
\AtBeginDocument{%
  }

\usepackage{tabularx}
\usepackage[table,usenames,dvipsnames]{xcolor}
\usepackage{stfloats}

\begin{document}

\title{InterA11y: Interactive Web Accessibility Error Detection with LLM Agents}

\author{Mingyuan Zhong}
\affiliation{%
  \institution{University of Washington}
  \city{Seattle}
  \state{Washington}
  \country{USA}
}
\email{myzhong@cs.washington.edu}

\author{Mengqi Shi}
\affiliation{%
  \institution{University of Washington}
  \city{Seattle}
  \state{Washington}
  \country{USA}
}
\email{shi21@uw.edu}

\author{Davin Win Kyi}
\affiliation{%
  \institution{University of Washington}
  \city{Seattle}
  \state{Washington}
  \country{USA}
}
\email{davin123@cs.washington.edu}

\author{Ajit Mallavarapu}
\affiliation{%
  \institution{Cornell University}
  \city{Ithaca}
  \state{New York}
  \country{USA}
}
\email{am3574@cornell.edu}

\author{Xia Chen}
\affiliation{%
  \institution{Carnegie Mellon University}
  \city{Pittsburgh}
  \state{PA}
  \country{USA}
}
\email{stanleyistool@gmail.com}

\author{James Fogarty}
\affiliation{%
  \institution{University of Washington}
  \city{Seattle}
  \state{Washington}
  \country{USA}
}
\email{jfogarty@cs.washington.edu}

\author{Jacob O. Wobbrock}
\affiliation{%
  \institution{University of Washington}
  \city{Seattle}
  \state{Washington}
  \country{USA}
}
\email{wobbrock@uw.edu}

\renewcommand{\shortauthors}{Zhong et al.}

\begin{abstract}
  Large Language Models (LLMs) have shown potential to expand the coverage of web accessibility evaluation. However, current LLM-based checkers neglect much of the interactive and dynamic web contexts, resulting in limited coverage, high false discovery rates, and inconsistent performance across models. We present \textit{InterA11y}, an agentic accessibility evaluation system that combines rubrics and interactive tools to support LLM agents in evaluating accessibility effectively. The rubrics decompose WCAG success criteria (SC) into testable aspects and guide agents in collecting and interpreting evidence. The interactive tools inspect or operate a webpage deterministically to collect evidence, such as focus appearances and error messages. We developed these rubrics and tools on 682 curated test cases grounded in established accessibility guidelines that target 10 commonly violated SC.
  We evaluated InterA11y on 197 expert-annotated cases sampled from 54 real webpages. Results show InterA11y achieves higher recall than the strongest evaluated LLM-driven baseline (79.0\% vs. 44.4\%), while maintaining comparable precision to best rule-based checkers (93.1\% vs. 90.3--97.6\%), with little variability between models. Further, when equipped with our interactive tools, general-purpose web agents detect more accessibility errors with fewer false positives and less token usage than without these tools.
  We release our code and data at \href{https://github.com/jszh/InterA11y}{\textcolor{NavyBlue}{this link}}.
\end{abstract}





\maketitle

\section{Introduction}
    \input{sections/10-introduction}

\section{Related Work}
    \input{sections/20-related-work}

\section{Evaluation of Current Checkers}
    \input{sections/30-current-state}

\section{The InterA11y System}
    \input{sections/40-intera11y}

\section{Expert Evaluation}
    \input{sections/50-expert-eval}

\section{Discussion}
    \input{sections/70-discussion}

\section{Conclusion}
    \input{sections/80-conclusion}


\section*{Ethics and Privacy Statement}

The expert study was reviewed by our University's Institutional Review Board (IRB) and received an exempt determination.
The implementation of InterA11y, including its prompts, tools, and harness, was supported by AI with extensive guidance from the authors.
The experiments and data analysis were supported by AI, with manual verification of data and results by the authors.

\bibliographystyle{ACM-Reference-Format}
\bibliography{reference}

\onecolumn
\appendix
\section{Details and Examples}\label{sec:app}
    \input{sections/99-app}

\end{document}

%% file: sections/10-introduction.tex
Identifying accessibility barriers in websites and web applications is essential to ensuring equitable web access.
The growing use of LLMs in web development makes the need for reliable accessibility evaluation more urgent~\cite{mowar2025codea11y,gurita2025llm}.
Currently, web accessibility evaluation requires substantial expertise~\cite{pereira2025evaluating} and manual efforts~\cite{velleman2014wcagem,webdevTestManual}, because
automated rule-based checkers provide limited coverage of guidelines~\cite{axe, wave} and relying on them alone can be risky~\cite{vigo2013benchmarking,calo2026measuring}.
LLM-based checkers have been proposed to address this coverage gap, especially for assessing errors that require semantic understanding of webpage structure and content~\cite{fathallah2025accessguru,he2025gena11y,lopezgil2025web,paterno2025how}.

Despite these potential benefits, current LLM-based checkers are limited in two respects: (1) lack of dynamic or interactive probing and (2) high variability across models.
First, current checkers examine static web contexts, such as HTML, JavaScript, and screenshots.
Accessibility evaluation under the  four Web Content Accessibility Guidelines (WCAG) principles---perceivable, operable, understandable, and robust---involves examining how webpages behave during user interaction, particularly for criteria related to keyboard navigation, focus management, and feedback~\cite{w3c_wcag2.2}.
In particular, ``operable'' targets the accessible interaction of web components but is largely left out by current LLM-based checkers, which have mostly targeted static content such as text alternatives and labels~\cite{aljedaani2026llm}.
Second, as we will show through our evaluation of current checkers, they are susceptible to different foundation models when tested on the same detection tasks, with false discovery rates (FDR) varying up to 26.5 percentage points across models. 
Such variability raises concerns about the reliability of LLM-based checkers in practical web development and testing workflows. Similarly, prior work has shown that general-purpose AI agents also face reliability challenges on several WCAG success criteria~\cite{santana2026limits}.

To improve LLMs' performance in web accessibility evaluation,
motivated by prior work where LLM agents are used to identify accessibility errors through interaction~\cite{zhong2026taskaudit}, 
we introduce \textbf{InterA11y}, an agentic accessibility evaluation system supported by \emph{interactive tools} and \emph{rubrics}.
The tools render and interact with a page programmatically, collecting evidence of markups and metadata, rendered appearance, and how it responds to keyboard, focus, and screen reader usage.
The rubrics support an LLM agent in interpreting this evidence and running additional tools in multiple turns to determine whether an aspect of a WCAG success criterion (SC) is met.
Our rubrics are informed by existing checkers (GenA11y~\cite{he2025gena11y}), guidelines~\cite{w3c_wcag2.2,w3c-act-rules}, and regulations~\cite{trusted_tester_5.1,etsi2021en301549}.
In this paper, we focus on 10 success criteria that are particularly relevant to interactive web contexts, covering all four WCAG principles, e.g., Non-text Content~(1.1.1), Focus Order~(2.4.3), Error Identification~(3.3.1), and Status Messages~(4.1.3).

To develop our system, we curated 682 test cases covering these 10 success criteria.
The test set includes 223 cases from the W3C's Accessibility Conformance Testing (ACT) test cases~\cite{w3c-act-rules} and 459 AI-generated cases independently validated by the authors.
We separately analyze the contributions of rubrics and agentic interactive tools using this test set.
Results show that adding InterA11y rubrics over GenA11y improved recall from 31.0\% to 64.2\%, while reducing false discovery rate (FDR) from 39.2\% to 31.1\%.
Adding interactive tool use further increased recall to 76.4\% and decreased FDR to 9.4\%---lower than many rule-based checkers.


We evaluated InterA11y with 197 expert-labeled cases sampled from 54 real webpages.
We compared against three types of baselines: (1) rule-based checkers, (2) existing LLM-based checkers, and (3) general-purpose web agents.
Results show that InterA11y achieves higher recall (79.0\%) than the strongest evaluated general-purpose web agent (36.0\%) and LLM-based checker (44.4\%).
InterA11y maintains comparable precision to rule-based checkers (93.1\% vs. 95.7\%), while general-purpose web agents had much lower precision ($\sim\!77\%$).
Further, adding InterA11y's interactive tools to the general-purpose web agents improved both their precision and recall while lowering token usage and execution time.

In summary, we make the following contributions in this paper:

\begin{enumerate}
    \item An empirical study of LLM-based web accessibility checkers and a characterization of their failure modes and their variability in detection performance.

    \item The InterA11y system, which combines rubrics and interactive tools to support reliable agentic evaluation of 10 WCAG success criteria across four accessibility principles.

    \item An evaluation of InterA11y, including controlled ablations on 682 curated test cases and comparisons with baseline techniques on 197 expert-labeled cases.
\end{enumerate}

%% file: sections/20-related-work.tex
Our work is grounded in web accessibility standards and informed by prior work in accessibility evaluation practices, rule-based, LLM-based, and agentic accessibility checkers.

\subsection{Web Accessibility Standards and Practices}
The Web Content Accessibility Guidelines (WCAG), developed and maintained by the World Wide Web Consortium (W3C), is the primary international standard for web accessibility~\cite{w3c_wcag2.2}. WCAG is organized around four core principles, known as the acronym ``POUR'': Perceivable, Operable, Understandable, and Robust. These principles are operationalized through success criteria, which are commonly mandated by governmental regulations~\cite{section508_2018, doj2024adatitle2, etsi2021en301549}.

To guide accessibility evaluation in practice, the W3C developed the WCAG Evaluation Methodology (WCAG-EM), which provides a structured process for assessing accessibility across entire websites~\cite{velleman2014wcagem}. 
However, manual evaluation remains resource-intensive, requiring trained evaluators and substantial time to assess compliance with WCAG success criteria~\cite{gu2026towards}. Many criteria also require contextual judgment~\cite{huq2023a11ydev} or examination of how users interact with web content~\cite{jordan2024information}, making comprehensive evaluation difficult to scale. The need for expert judgment, contextual interpretation, and time-intensive manual evaluation motivates the development of automated approaches that can reduce evaluation effort while supporting a broader range of accessibility requirements.

\subsection{Rule-Based Checkers and Their Limitations}
Automated accessibility checkers reduce the effort required for manual evaluation by identifying potential WCAG violations. Existing approaches can be broadly divided into static checkers, which primarily inspect webpage structure and properties, and dynamic checkers, which additionally execute interactions to examine runtime behavior. Both approaches have improved the scalability of accessibility evaluation, but their coverage remains limited.

\subsubsection{Static Checkers}
Static checkers such as axe~\cite{axe}, WAVE~\cite{wave}, and HTML validators~\cite{htmlcs,pavel2018verifying} primarily inspect webpage markup, DOM structures, and element properties to identify accessibility violations~\cite{taeb2024axnav}. They are fast, easy to integrate into development workflows, and capable of evaluating large numbers of webpages automatically~\cite{vigo2013benchmarking}.
Some checkers are specifically engineered to limit false positives~\cite{axe}.

Static checkers are constrained by predefined heuristics and static contexts, limiting their coverage in accessibility evaluation~\cite{zhong2025screenaudit, he2025gena11y}.
Fischer et al. systematically analyzed widely-used automated accessibility checkers 
and found that they collectively covered only one sixth of WCAG success criteria~\cite{fischer2025coverage}.
Similarly, He et al. reported that WAVE covers only 13 of 86 WCAG success criteria~\cite{he2025gena11y}. 
Prior research has also documented the tradeoff between their false positives and coverage~\cite{vigo2013benchmarking}. 

\subsubsection{Dynamic Checkers}
To extend automated evaluation into interactive aspects, dynamic checkers interact with webpages or alter properties to inspect dynamic behavior.
By observing how interfaces respond to programmatic actions, these tools can identify accessibility issues that may only become apparent during interaction, such as keyboard navigation failures~\cite{chiou2021keyboard, chiou2023dialog,chiou2023bagel,bhagat2026satya}, focus indicator visibility~\cite{bhagat2026satya}, and missing feedback~\cite{ko2011feedlack}.
Another line of work modifies viewport widths to inspect layout~\cite{walsh2017redecheck} and reflow~\cite{chiou2024reflow} failures.

However, interaction alone does not overcome the limitations of predefined rules. Rule-based dynamic checkers can observe interface behavior, but their assessments are constrained by the conditions encoded in their detection logic and lack of language understanding. Similar limitations have been observed in mobile accessibility evaluation, where rule-based checkers miss errors involving screen reader feedback~\cite{zhong2025screenaudit} and interactive functionalities~\cite{zhong2026taskaudit}.

Static and dynamic rule-based checkers have improved the scalability of automated accessibility evaluation, but both face limitations in coverage and contextual interpretation. Addressing these limitations motivates approaches that combine semantic reasoning with evidence collected through webpage interactions.

\subsection{LLM-Based and Agentic Checkers}
Recent research has explored the use of Large Language Models (LLMs) to extend automated accessibility evaluation beyond predefined rules~\cite{aljedaani2026llm}. For example, GenA11y~\cite{he2025gena11y}, AccessGuru~\cite{fathallah2025accessguru}, and López-Gil and Pereira~\cite{lopezgil2025web,lopezgil2026empirical} use LLMs to identify web accessibility issues that require semantic understanding, while Gu et al.~\cite{gu2026towards} investigate the use of multimodal LLMs as copilots for web accessibility auditing. Prior research demonstrates how LLM-based techniques can extend accessibility evaluation coverage.

Current LLM-based checkers face similar limitations as static rule-based checkers, in that they mostly ingest static code and screenshots without consideration of dynamic contexts.
Recent research has started to explore agentic accessibility evaluation.
Mohanbabu et al. observed significantly degraded performance when computer-use agents performed everyday computer tasks using assistive technologies~\cite{mohanbabu2026a11ycua}.
Santana et al. further demonstrated that successful web task execution and accurate accessibility evaluation are distinct capabilities, and general-purpose web agents remain unreliable at the latter~\cite{santana2026limits}.
In mobile apps, TaskAudit built a custom agent to operate screen readers and detect accessibility errors, but are still susceptible to false positives and unreliable task execution~\cite{zhong2026taskaudit}. 

However, dynamic and agentic testing of webpages are less studied and understood.
In this paper, we first conduct experiments to understand current challenges, then propose techniques to support more reliable agentic accessibility assessment.

%% file: sections/30-current-state.tex
We begin by examining current rule-based and LLM-based accessibility checkers' performance on test cases specifically designed to cover various dimensions of WCAG success criteria.
Specifically, we selected 10 representative success criteria based on whether they require manual assessments in practice~\cite{deque2024coverage} and whether prior work has identified a need for dynamic testing.
Table~\ref{tab:sc} provides a general description of the selected criteria.
For these criteria, we curated a set of 682 test cases that are designed to assess checkers' conformance to WCAG.

\begin{table}[t]
    \centering
    \caption{Selected WCAG success criteria, with ranks by the number of cases requiring manual testing~\cite{deque2024coverage}, and relevant references motivating the need for dynamic checking.}
    \label{tab:sc}
    \small
    \setlength{\tabcolsep}{3pt}
    \begin{tabularx}{\columnwidth}{@{}lXlrl@{}}
    \toprule
    \textbf{SC} & \textbf{Title} & \textbf{Principle} &
    \textbf{Rank} & \textbf{Reference} \\
    \midrule
    \href{https://www.w3.org/WAI/WCAG22/Understanding/non-text-content.html}{1.1.1}
    & Non-text Content & Perceivable & \#6
    & \cite{zhong2025caption} \\
    
    \href{https://www.w3.org/WAI/WCAG22/Understanding/use-of-color.html}{1.4.1}
    & Use of Color & Perceivable & \#11
    & \cite{chiu2026megacoin,hayashi2026color} \\
    
    \href{https://www.w3.org/WAI/WCAG22/Understanding/content-on-hover-or-focus.html}{1.4.13}
    & Content on Hover or Focus & Perceivable & \#21
    & \cite{w3c_wcag2.2} \\
    
    \href{https://www.w3.org/WAI/WCAG22/Understanding/keyboard.html}{2.1.1}
    & Keyboard & Operable & \#5
    & \cite{chiou2021keyboard,chiou2023bagel,bhagat2026satya} \\
    
    \href{https://www.w3.org/WAI/WCAG22/Understanding/no-keyboard-trap.html}{2.1.2}
    & No Keyboard Trap & Operable & \#29
    & \cite{he2025gena11y,chiou2023dialog} \\
    
    \href{https://www.w3.org/WAI/WCAG22/Understanding/focus-order.html}{2.4.3}
    & Focus Order & Operable & \#4
    & \cite{he2025gena11y,chiou2023bagel,gu2025inclusive} \\
    
    \href{https://www.w3.org/WAI/WCAG22/Understanding/link-purpose-in-context.html}{2.4.4}
    & Link Purpose (In Context) & Operable & \#16
    & \cite{kumar2025pdf,lopezgil2025web} \\
    
    \href{https://www.w3.org/WAI/WCAG22/Understanding/focus-visible.html}{2.4.7}
    & Focus Visible & Operable & \#7
    & \cite{chiou2023bagel,bhagat2026satya} \\
    
    \href{https://www.w3.org/WAI/WCAG22/Understanding/error-identification.html}{3.3.1}
    & Error Identification & Understandable & \#22
    & \cite{w3c_wcag2.2} \\
    
    \href{https://www.w3.org/WAI/WCAG22/Understanding/status-messages.html}{4.1.3}
    & Status Messages & Robust & \#17
    & \cite{w3c_wcag2.2} \\
    \bottomrule
    \end{tabularx}
\end{table}

\subsection{Curating a Conformance Test Set}\label{sec:test-set}

We curate a test set for evaluating accessibility checkers' conformance to WCAG success criteria by combining relevant ACT test cases and LLM-generated additional test cases.
Our conformance test set includes 682 cases in total (223 ACT and 459 LLM-generated), covering 72 rules in 10 success criteria.
Each criterion has between 25 and 174 test cases ($M=68.2$, $SD=50.6$).
316 of the 682 cases are positive (46.3\%) and 366 are negative.

\subsubsection{ACT Test Cases}
We first included ACT test cases relevant to the selected criteria.
The ACT rules are developed by the W3C to ``test conformance with Web Content Accessibility Guidelines (WCAG) 2.2, ... and other accessibility practices''~\cite{w3c-act-rules}.
Although passing these rules does not guarantee WCAG compliance, the rules represent an industry consensus on how WCAG should be implemented~\cite{w3c-act-rules-about}.
In total, there are 223 ACT test cases covering 14 ACT rules on 6 of the 10 selected criteria.

\subsubsection{LLM-generated Test Cases}
The ACT only provides a partial coverage of the select criteria, with especially limited coverage on the interactive aspects.
We constructed an expansion set to the ACT test cases, following the same design principles: each case should target one aspect of a criterion, be a minimal representation of the requirement, and be easily consumable by testing tools.
Two of the authors examined each of the selected criteria, including its official documentation~\cite{w3c_wcag2.2}, ACT rules~\cite{w3c-act-rules}, assessment standards from governmental agencies~\cite{section508_2018, etsi2021en301549}, and industry best practices~\cite{webdevTestManual}.
From these standards and guidelines, the two authors individually identified a list of additional rules needed for each criterion, then discussed to reach a final consensus list.
In total, the authors identified 58 additional rules across 9 success criteria.
While these are unlikely to cover all situations where a person with disability might encounter a barrier, we believe they offer a breadth of coverage that reflects the intent of current standards.

We then used Claude Code~\cite{claude_code} with Opus 4.6 to generate passing and failing test cases for these rules.
For each rule, the coding agent is supplied with success criteria documentation and relevant excerpts from standards and guidelines.
It is prompted to generate a variety of instances where these rules are applicable (e.g., by varying content type, UI component, language, technical implementation).

Four authors independently annotated the generated cases, with each case receiving two annotations.
Each annotator was asked to assess, using necessary assistive tools or by examining source code, whether each test case's label was correct and whether any labels leaked into the test case itself.
Annotators received guidelines relevant to each test case's success criterion.
Each annotator first practiced annotations on 20 ACT test cases until they were familiar with the interface and instructions.
We excluded any test case where at least one annotator rejected and removed leaked labels from any flagged test cases.

Claude Code initially generated 511 test cases for the 58 rules.
The authors each reviewed between 250 and 260 test cases, resulting in 1022 annotations.
Annotators agreed on 89.8\% of the cases, with Cohen's $\kappa=0.814$ indicating strong agreement.
We excluded 52 of the 511 cases where at least one author rejected, resulting in 459 final expansion cases.


\subsection{Experiment Setup}
In this experiment, we ran rule-based checkers and LLM-based checkers from publicly available implementations.
Experiments ran in isolated Google Chrome browsers on a Rocky Linux 10.2 server, using $1280\times900$~px  viewports, except when resized by tools.

\subsubsection{Rule-Based Checkers}
We include 11 static and dynamic rule-based checkers: Alfa~\cite{alfa}, ASLint~\cite{aslint}, axe~\cite{axe}, BAGEL\footnote{BAGEL was implemented according to its technical description as its source code was not available.}~\cite{chiou2023bagel}, Editoria11y~\cite{Editoria11y}, HTML\_CodeSniffer~\cite{htmlcs}, IBM Equal Access~\cite{ibm-equal-access-act}, LYNX~\cite{winn2026lynx}, Nu Html Checker~\cite{nu_checker}, QualWeb~\cite{qualweb}, and Testaro~\cite{testaro}.
Following Pool's finding that ensemble testing with rules from different checkers improves testing performance~\cite{pool2023metatesting}, we also fit and compute the best rule ensemble\footnote{The rule ensemble is fitted using mixed integer programming~\cite{wolsey1998integer} to maximize F1 score, based on rule-level combinations of the 11 checkers.} on our conformance test set, approximating a best possible rule-based checker.

\subsubsection{LLM-Based Checkers}
We include three LLM-based checkers from prior work:
GenA11y~\cite{he2025gena11y}, which extracts page contents per success criterion and judges accessibility with an LLM; AccessGuru~\cite{fathallah2025accessguru}, which combines axe and an LLM-based semantic detector;
and López-Gil and Pereira's technique~\cite{lopezgil2026empirical}, which prompts an LLM with criterion description and explanations.
We include these three because of their broad coverage, public availability, and complementary approaches.
While the public implementations of all three checkers use older models (e.g., GPT-4o, Gemini 2.0 Flash), we test them using recent models for fairness and consistency.
To support the evaluation of variability across foundation models, we test with four models: Claude Haiku 5.5, Gemini 3.7 Flash, GLM 5.3 Flash, and GPT-6 Luna, all with reasoning set to ``high.''

\begin{figure}[b]
\centering
 \includegraphics[width=0.94\linewidth]{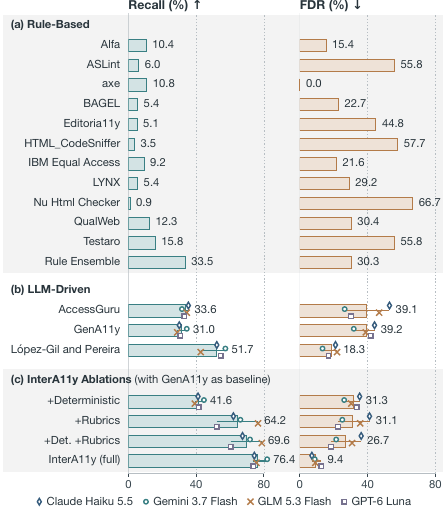}
 \vspace{-4px}
  \caption{Mean recall and FDR of accessibility checkers on the conformance test set. Error bars show ranges across models.}
  \Description{}
  \label{fig:ex1}
\end{figure}

\begin{figure*}[b]
\centering
 \includegraphics[width=0.93\linewidth]{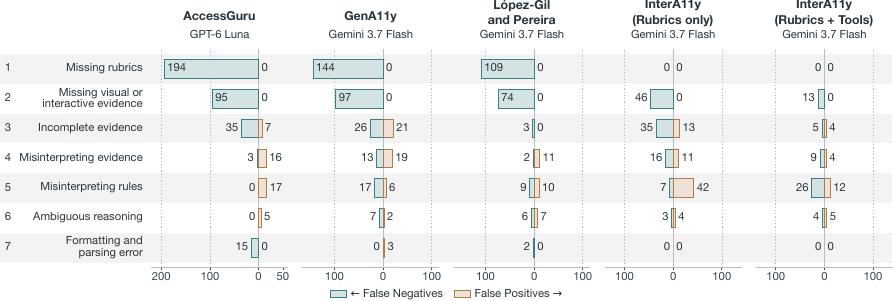}
 \vspace{-4px}
  \caption{Error reasons of accessibility checkers using their best evaluated models. Each error can map to multiple reasons. 
  }
  \Description{}
  \label{fig:ex1-error}
\end{figure*}

\subsection{Results}\label{sec:exp1-results}
\subsubsection{Recall and False Discovery Rate}
Figures~\ref{fig:ex1}a and~\ref{fig:ex1}b compare the recall and false discovery rate ($FDR=1-precision$) for all checkers evaluated.
The fitted rule ensemble~\cite{pool2023metatesting} includes 65 unique rules drawn from all eleven rule-based checkers.

Rule-based checkers consistently scored lower recall on this set that requires dynamic contexts.
Surprisingly, many rule-based checkers had high false detection rates, with a mean FDR of 36.4\% and only axe achieving a 0\% FDR, consistent with earlier benchmarks that found higher completeness to come with lower correctness~\cite{vigo2013benchmarking}.
The rule ensemble achieved best recall (33.5\%) among rule-based checkers, but still had high FDR (30.3\%).
One possible explanation is that our test set oversamples edge cases and may not reflect real-world accessibility error distributions, but serves as a stress test.

All three LLM-based checkers achieved higher recall than any single rule-based checker, showing LLMs' strong potential in web accessibility assessment.
The rule ensemble approaches or surpasses several models tested.
LLM-based checkers still have limited coverage in these criteria, with recall around 30--50\%.
We also see large performance variability across foundation models.
For example, the largest range of recall is 14.6 percentage points (López-Gil and Pereira) and the largest range of FDR is 26.5 percentage points (AccessGuru).
A model can perform best in one system, but worst in another.
Such sensitivity to models can make it difficult to apply LLM-based checkers in practice.

\subsubsection{LLM-Based Checker Error Analysis}
For the model with best F1 in each LLM-based checker, we analyzed reasons for its false positives and false negatives.
Figure~\ref{fig:ex1-error} breaks down the causes, which can be broadly categorized into three types: \textit{rules and rubrics} (\#1, \#5), \textit{evidence} (\#2, \#3, \#4), and \textit{model output} (\#6, \#7).
Across all three checkers, most false negatives were due to missing rubrics in prompts and missing or incomplete evidence.
Most false positives were due to incomplete evidence and misinterpreting evidence or rules.
Model errors represented a small portion of all errors.

%% file: sections/40-intera11y.tex
We design the InterA11y system to address the two main error sources identified in Section~\ref{sec:exp1-results}: rubrics and evidence.
To quantify and compare the improvements brought by new rubrics and evidence collection techniques, we develop InterA11y based on the closest baseline, GenA11y~\cite{he2025gena11y}, which analyzes accessibility by success criterion and already collects basic evidence, e.g., HTML contents, CSS properties, and screenshots.

InterA11y is designed to be an agentic system: the agent invokes \textit{interactive tools} in multiple turns as needed, following the guidance of \textit{rubrics}.
The InterA11y harness supports this process, where it manages the collected evidence, LLM contexts, and browser resources.
We re-use GenA11y's rubrics where applicable but develop our own harness, tools, and rubrics.


\subsection{InterA11y System Design}
\begin{figure*}[t]
\centering
 \includegraphics[width=\linewidth]{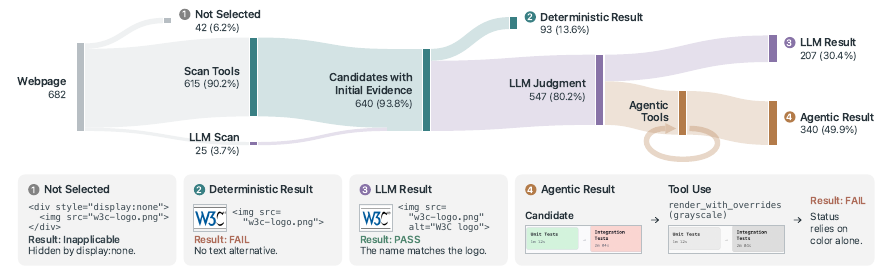}
 \vspace{-15px}
  \caption{Illustration of the InterA11y data pipeline from a Gemini 3.7 Flash run on the 682-item conformance test set. Numbers under each stage indicate the test cases reaching that stage. (1)--(4) show examples of four possible outcomes from the pipeline.}
  \Description{}
  \label{fig:system}
\end{figure*}

Figure~\ref{fig:system} illustrates how webpages are assessed through the InterA11y system pipeline.
For each success criterion, the harness applies relevant \textit{scan tools} to collect initial evidence and conducts an LLM scan\footnote{The LLM scan is not modified from the GenA11y implementation~\cite{he2025gena11y}.} to identify candidate elements.
It combines candidates identified through heuristics and the LLM scan, then applies deterministic checks for cases with conclusive measurements (i.e., similar to rule-based checkers).
Then, an LLM guided by rubrics evaluates the evidence and collects new evidence using \textit{agentic tools} as needed.
Once sufficient evidence is collected, the model generates a final verdict for each candidate.

\subsubsection{Interactive Tools}
Table~\ref{tab:tools} provides an overview of InterA11y's interactive tools.
The tools can be broadly divided into five categories: 
    (1) \textit{inspection} of programmatic page information, 
    (2) \textit{capture} of rendered visual information, 
    (3) \textit{measurement} of rendered properties, such as contrast, 
    (4) \textit{interaction} of inputs and controls, and
    (5) \textit{monitoring} of page responses to interaction.
Although GenA11y~\cite{he2025gena11y} included several scan tools, we found them to be insufficient in evidence collection based on our previous analysis and therefore did not adopt them in our system.
In our implementation, we pay close attention to reproducing assistive technology output realistically (e.g., screen reader).
To this end, we integrate an open-source virtual screen reader built by Guidepup,\footnote{Virtual Screen Reader: \url{https://www.guidepup.dev/}} which could be replaced by real assistive technology streams in a production system.
Tools are also engineered to accommodate loading times and support robust re-tries.
In total, we built 46 scan and agentic tools, including 24 in interaction and monitoring.
A list of all tools is available in Appendix~\ref{sec:app}.

\begin{table}[t]
    \centering
    \caption{InterA11y's scan tools and agentic tools by category.}
    \label{tab:tools}
    \small
    \begin{tabular}{lrr}
    \toprule
    \textbf{Category} & \textbf{Scan tools} & \textbf{Agentic tools} \\
    \midrule
    Inspection  & 4  & 2 \\
    Capture     & 3  & 3 \\
    Measurement & 6  & 4 \\
    Interaction & 18 & 2 \\
    Monitoring  & 2  & 2 \\
    \bottomrule
    \end{tabular}
\end{table}

\subsubsection{Rubrics}
InterA11y follows rubrics in its analysis of each candidate to reason about available evidence, call necessary tools, and make a decision.
Table~\ref{tab:rubrics} provides a summary of InterA11y's rubrics.
In authoring the rubrics, we considered WCAG Understanding documents,\footnote{WCAG 2.2 Understanding Docs: \url{https://www.w3.org/WAI/WCAG22/Understanding/}} Techniques,\footnote{Techniques for WCAG 2.2: \url{https://w3c.github.io/wcag/techniques/}} ACT rules~\cite{w3c-act-rules}, and U.S.~\cite{trusted_tester_5.1} and European~\cite{etsi2021en301549} testing requirements.
GenA11y~\cite{he2025gena11y} provides a validated starting point for our rubrics development.
Of the 35 rubric items included in InterA11y, 11 were originally developed in GenA11y, covering 4 success criteria.
However, these rubrics do not include aspects related to interaction, feedback, and tool use.
Therefore, we expand them to cover all 10 selected criteria by developing an additional 24 rubric items.
Each rubric item decides one aspect of a criterion, specifying evidence required and decision method based on WCAG's documentations.

\begin{table}[t]
\centering
    \caption{InterA11y's rubrics by category and by source.}
    \label{tab:rubrics}
    \small
    \begin{tabular}{llrrr}
    \toprule
    \textbf{Category} & \textbf{WCAG SC} & \textbf{GenA11y} & \textbf{InterA11y} & \textbf{Total} \\
    \midrule
    Content alternatives & 1.1.1 & 5 & 4 & 9 \\
    Link purpose & 2.4.4 & 3 & 1 & 4 \\
    Keyboard and focus &
    \begin{tabular}[t]{@{}l@{}}
    2.1.1, 2.1.2, \\
    2.4.3, 2.4.7
    \end{tabular} & 0 & 11 & 11 \\
    Cues and feedback &
    \begin{tabular}[t]{@{}l@{}}
    1.4.1, 1.4.13, \\
    3.3.1, 4.1.3
    \end{tabular} & 3 & 8 & 11 \\
    \bottomrule
    \end{tabular}
\end{table}

\subsubsection{Illustrative Examples}
Figure~\ref{fig:system} shows four exemplar webpage inputs processed by the InterA11y pipeline.
Example (1) shows an element excluded by the harness, because it is hidden.
Example (2) shows an image without an alt text and is deterministically decided to fail.
Example (3) shows an image with proper alt text and is decided by the LLM based on its screenshot without using agentic tools.
Example (4) shows UI elements that use color to represent meaning. The rubric for SC 1.4.1 includes the language: ``A graphic fails when categories or states are distinguished only by colour...'' with available tools listed.
The agent thus uses the tool to view a grayscale version of the elements, then reaches the ``fail'' verdict.
Additional illustrative examples are available in Appendix~\ref{sec:app}.

\subsection{Ablation Study}
To understand the contribution of rubrics and interactive tools on InterA11y's performance, we conducted an experiment with five ablations: (1) original GenA11y~\cite{he2025gena11y}, (2) GenA11y + deterministic checks, replacing GenA11y's heuristic tools, (3) GenA11y + InterA11y's rubrics, (4) GenA11y + deterministic checks + InterA11y's rubrics, and (5) the full InterA11y system with deterministic checks, rubrics, and agentic tool execution.
The experiment was run on the conformance test set described in Section~\ref{sec:test-set}, using the same LLMs and configurations.
Statistical analyses were conducted on the case level with two-sided paired permutation tests~\cite{winkler2014permutation} with Holm--Bonferroni correction~\cite{holm_bonferroni}.

\subsubsection{Recall and False Discovery Rate}
Figure~\ref{fig:ex1}c shows the performance of the InterA11y ablations, while GenA11y's performance is shown in Figure~\ref{fig:ex1}b.
While adding deterministic checks improves average recall and FDR by around 10\%, the main improvements are from rubrics and interactive tool use.
Adding InterA11y's rubrics improves mean recall from 31.0\% to 64.2\% ($p<.001$) and reduces mean false discovery rate (FDR) from 39.2\% to 31.1\% ($p=.057$).
Adding agentic tool use increased recall from 69.6\% to 76.4\% ($p=.18$) and decreased FDR from 26.7\% to 9.4\% ($p<.001$).
Adding evidence collected by tools also made the system robust to model differences, reducing the range of recall across four foundation models from 24.4 in rubrics only to 8.2 percentage points ($p=.001$) in the full system and the range of FDR from 18.6 to 5.6 points ($p=.001$), best among all LLM-based checkers evaluated.

\begin{figure}[t]
\centering
 \includegraphics[width=0.95\linewidth]{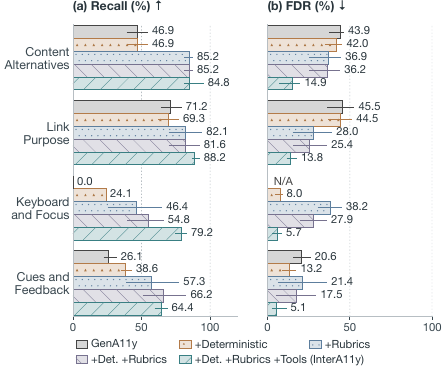}
 \vspace{-4px}
  \caption{Mean recall and FDR across accessibility error categories in the ablation.  Error bars show ranges across models.}
  \Description{}
  \label{fig:ablation}
\end{figure}

Figure~\ref{fig:ablation} shows ablation performance by rubrics category according to Table~\ref{tab:rubrics}.
Both InterA11y's rubrics and interactive tools are especially useful for improving recall of ``keyboard and focus,'' while rubrics alone are useful in improving recall of ``content alternatives'' and ``cues and feedback.''
Across all categories, adding interactive tools reduces FDR over rubrics and GenA11y.

\subsubsection{Error Analysis}
Following the same approach as in Section~\ref{sec:exp1-results}, we analyzed the error reasons of InterA11y's ablations.
Figure~\ref{fig:ex1-error} compares InterA11y's error reasons with existing LLM-based checkers.
Results show that our rubrics fill a significant gap left by current checkers (\#1).
Agentic tool use further reduces the errors caused by missing, incomplete, or misinterpreted evidence (\#2, \#3, \#4) from 121 to 35, a 71\% reduction.
Interestingly, adding tool use removed 30 false positives caused by misinterpreting rules (\#5), but introduced 19 false negatives.
Overall, the full InterA11y system provides significant improvements to existing LLM-based checkers thanks to improved rubrics coverage and agentic tool use.





%% file: sections/50-expert-eval.tex
Because our initial development and analyses focused on minimal test cases that are less representative of real-world webpages, we conducted an evaluation on 197 expert-annotated cases from 54 real webpages.
Informed by Santana et al.'s recent work on AI agents' ability to evaluate accessibility~\cite{santana2026limits}, we compared InterA11y with rule-based checkers, LLM-based checkers, and two general-purpose agents: Browser Use~\cite{browser_use2024} and AgentLab~\cite{chezelles2025browsergym}.
These web agents are equipped with generic tools designed for browser-related tasks, such as input, scroll, and form controls.
We prompted each agent using the corresponding InterA11y rubrics.

As a follow-up, we introduced InterA11y's tools to the general-purpose agents and evaluated their performance differences.
We also compared token usage and execution time across all checkers.

\subsection{Dataset Construction}
To construct a diverse dataset covering major uses of the web,
we sampled 54 webpages from 48 websites in 14 areas, e.g., retail, education, finance, social networking, and public record services.
A list of all websites is available in Appendix~\ref{sec:app}.
For reproducibility, we saved the webpages' code and resources.
We randomly sampled 300 elements from these saved webpages.
An author manually went through the elements and filtered out obviously out-of-scope ones, including videos, elements that are neither visible nor focusable, and elements that require dynamic server resources.
In total, we included 211 elements for annotation.

Two experts in accessibility, each with over five years of experience developing or researching assistive technologies, were recruited for this annotation.
For each element, a brief description of a success criterion was shown, and the target element was highlighted on the webpage. Each annotator was asked to inspect it regarding the provided criterion.
The experts each annotated 103 and 108 elements, respectively.
Two authors with prior accessibility research experience  separately annotated these 211 elements.
In total, each element received annotations from one expert and one author.
We compared these annotations and found 14 disagreements.
These cases were sent to the other expert for review, and we recorded the majority label.
In addition, 14 other cases were found to be unreliable (e.g., webpage contents changed) and were excluded.
The final expert-annotated dataset contained 197 unique elements, with 107 positives (54.3\%) and 90 negatives.

\subsection{Experiment Setup}
We used the same configurations for rule-based and LLM-based checkers tested in Section~\ref{sec:exp1-results}, but excluded the checker by López-Gil and Pereira~\cite{lopezgil2026empirical}, because (1) it was designed for small test cases and did not generate element-level findings; and (2) it sent full HTML in its prompts, exceeding token limits in at least 17 cases.

For Browser Use~\cite{browser_use2024} and AgentLab~\cite{chezelles2025browsergym}, we used their default configuration.
We provided relevant InterA11y rubrics as their initial user message (task).
For the conditions with InterA11y tools, we registered our tools using APIs recommended in their documentations.
Otherwise, we kept their configurations the same.
Statistical analyses were conducted on the case level with two-sided paired permutation tests~\cite{winkler2014permutation} with Holm--Bonferroni correction~\cite{holm_bonferroni}.

\subsection{Results}

\subsubsection{Recall and False Discovery Rate}
\begin{figure}[t]
\centering
 \includegraphics[width=0.95\linewidth]{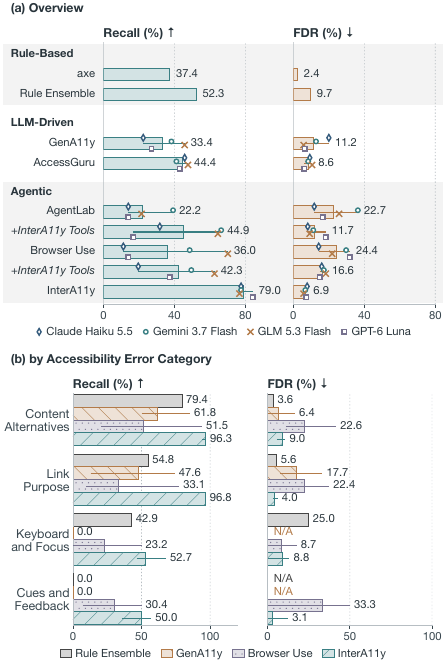}
 \vspace{-4px}
  \caption{Mean recall and FDR of accessibility checkers on expert-labeled elements. Error bars show ranges across foundation models.}
  \Description{}
  \label{fig:ex2}
\end{figure}

Figure~\ref{fig:ex2} shows the performance comparison between systems evaluated.
For brevity, we only show the best rule-based checker by F1 (axe) and rule ensemble.
On these real-world websites, we observe comparable recall between axe and LLM-based checkers, with LLM-checkers averaging 3--4 times higher FDR.
Similar to Section~\ref{sec:exp1-results}, we still see variation between models, with GenA11y's recall ranging 23.4 percentage points (pp) and FDR ranging 14.2 pp.
Surprisingly, general-purpose web agents, with their many tools targeted at interaction, do not achieve better results on average.
Their FDRs are especially high, more than twice that of LLM-based checkers on average.
Prior work has shown the unreliability of tool-use preferences in LLMs~\cite{faghih2025tool} and having too many tools may have exacerbated this problem.

InterA11y consistently achieves 76.6--84.1\% recall across models, with only a 2.0 percentage point FDR range between 5.7\% and 7.8\%.
Its recall is higher than all other checkers tested (all $p<.05$) and FDR is similar to axe (2.4\%, $p=.89$) and rule ensemble (9.7\%, $p=.92$).
Breaking down performance by category, InterA11y achieves almost perfect recall and high reliability for ``content alternatives'' and ``link purpose,'' while identifying 50\% ``cues and feedback'' issues when current checkers could detect almost none.
In terms of ``keyboard and focus,'' LYNX~\cite{winn2026lynx} and BAGEL~\cite{chiou2023bagel} rules contribute to recall of 42.9\%, but also a relatively high FDR of 25\%.
InterA11y detects 52.7\% of errors at an FDR of only 8.8\%.
Across all categories, InterA11y maintains an average FDR of less than 10\%.

\subsubsection{General-Purpose Web Agents}
\begin{figure}[t]
\centering
 \includegraphics[width=0.95\linewidth]{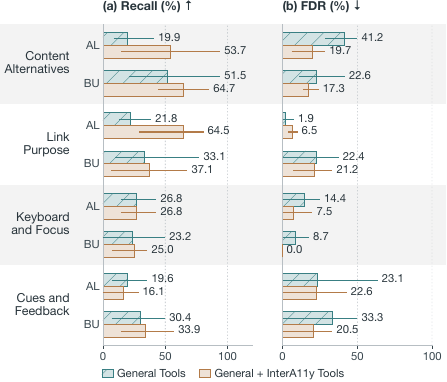}
 \vspace{0px}
  \caption{Mean recall and FDR of general-purpose web agents without and with InterA11y interactive tools, by accessibility error categories. Error bars show ranges across models. AL = AgentLab~\cite{chezelles2025browsergym}. BU = Browser Use~\cite{browser_use2024}.}
  \Description{}
  \label{fig:ex2-general-agents}
\end{figure}

To further understand the effectiveness of InterA11y's tools, we supply them to general-purpose agents.
On average, recall increases from 22.2\% to 44.9\% with AgentLab and from 36.0\% to 42.3\% with Browser Use.
FDR decreases from 22.7\% to 11.7\% with AgentLab and from 24.4\% to 16.6\% with Browser Use.
Figure~\ref{fig:ex2-general-agents} shows a breakdown by accessibility error category.
In most categories, introducing InterA11y tools improves or maintains recall while reducing FDR.
This is especially prominent for content alternatives and link purpose in AgentLab.

\subsubsection{Token Usage, Cost, Execution Time}
We analyze the cost implications of InterA11y in the context of LLM-based and general-purpose agents.
Averaged across models, InterA11y spends 31--81\% more tokens per case than LLM-based checkers, while improving recall by over 30 pp and reducing FDR.
InterA11y spends 14--38\% fewer tokens than general-purpose agents.
On average, using the cheapest model with InterA11y (GPT-6 Luna) costs \$0.0094 per case.
As a comparison, the cheapest model with an LLM-based checker (Claude Haiku 5.5 with AccessGuru) costs \$0.0030 per case.
InterA11y is the fastest agentic checker tested, taking 109 seconds per case on average.

\begin{figure}[t]
\centering
 \includegraphics[width=0.95\linewidth]{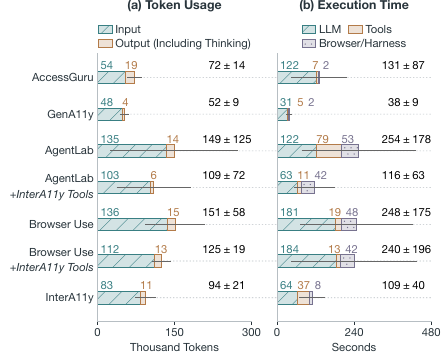}
 \vspace{-10px}
  \caption{Token usage and execution time per case, averaged across all models. Error bars show standard deviation.}
  \Description{}
  \label{fig:ex2-tokens}
\end{figure}

When both general-purpose agents are equipped with InterA11y tools, their token usage and completion time decrease on average.

%% file: sections/70-discussion.tex
InterA11y presents a novel approach of integrating deterministic tool use and LLM-based context understanding to identify challenging dynamic accessibility problems on the web.
Our tool design exposes browser interactions that are relevant to accessibility evaluation, echoing prior findings that web agent interfaces affect effectiveness and efficiency~\cite{steiner2026interfaces,yang2025agentoccam}.
InterA11y also parallels recent work in agentic web testing~\cite{teoh2026webtestpilot}, translating high-level requirements into evaluation procedures and grounding judgment in observed content and behavior.

By combining guideline-grounded rubrics and interactive tools, InterA11y provides an agentic solution to reducing false positives, enhancing reliability, and expanding coverage.
Our experiments further show that these interactive tools provide reusable evaluation capabilities across different web agents.

Automating these tasks frees up repetitive human efforts that currently require extensive training~\cite{trusted_tester_5.1} and time~\cite{pereira2025evaluating}, saving human expertise for more impactful tasks.
The improvements in false discovery rate and model variability can open up practical applications, such as integrating into CI workflows.
Prior work has also shown benefits of sharing agent-collected mobile accessibility evidence to testers~\cite{taeb2024axnav}.
Similarly, developer toolkits can present interactive evidence collected by InterA11y to support bug reporting, triaging, and reproduction of web accessibility issues.

Concurrent work by Mishra et al.~\cite{mishra2026agentic} similarly uses agents and browser tools, including interaction probes.
InterA11y instead organizes tool use around finer-grained rubrics.
While their evaluation focuses on page-level agreement, InterA11y systematically evaluates agentic tool use's contributions to element-level reliability, model variability, and their benefits to general-purpose web agents.



\section{Limitations and Future Work}

Due to limited availability of high-quality validated web accessibility datasets, we constructed our own conformance test set, which supported both our development and ablation study.
While the ablation still provided useful insights into each system component, future work should explore creating high-quality dynamic web accessibility testing datasets.

Our expert-annotated webpages were collected from real websites, but they are still simplified as offline replications.
Future work should explore and evaluate dynamic accessibility testing that covers server-side and authenticated functionalities.

%% file: sections/80-conclusion.tex
In this paper, we constructed a conformance test set designed for dynamic accessibility errors to characterize the main challenges to current LLM-based checkers, particularly in rubrics and evidence.
We then proposed InterA11y, an agentic accessibility checker that combines rubrics and interactive tools to support reliable and accurate detection.
InterA11y contributes a viable pathway for detecting dynamic web accessibility errors and ensuring equitable access.
Our results motivate the development of reusable tools grounded in guidelines to expand reliable web accessibility testing.

%% file: sections/99-app.tex
\begin{figure*}[t]
\centering
 \includegraphics[width=0.98\linewidth]{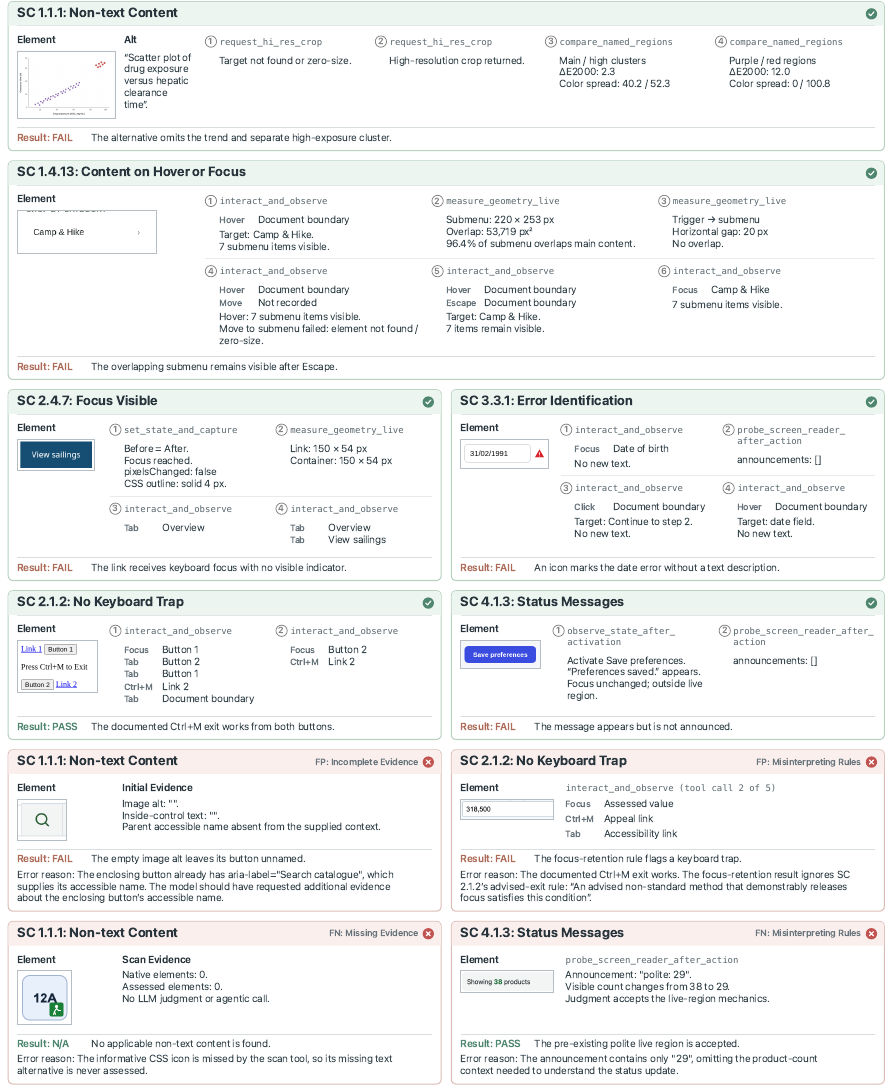}
 \vspace{-4px}
  \caption{Illustrative examples of InterA11y's correct and incorrect judgments, showing tool call traces and error reasons.}
  \Description{}
  \label{fig:trace-examples}
\end{figure*}

We provide additional details of InterA11y and expert study, as well as examples of InterA11y's execution.
Table~\ref{tab:intera11y-tools} shows a list of all scan tools and agentic tools and their corresponding success criteria.
Table~\ref{tab:website-areas} shows a list of all websites examined by experts.
Figure~\ref{fig:trace-examples} shows examples of InterA11y's correct and incorrect judgments with tool call traces and error reasons.

\newcommand{\iatool}[1]{\texttt{#1}}
\newcommand{\iatoollist}[1]{\begingroup\raggedright\setlength{\parindent}{0pt}\let\\\par #1\par\endgroup}
\newcommand{\iatwoscans}[2]{%
  \begin{minipage}[t]{\dimexpr.5\linewidth-2pt\relax}%
    \iatoollist{#1}%
  \end{minipage}\hspace{4pt}%
  \begin{minipage}[t]{\dimexpr.5\linewidth-2pt\relax}%
    \iatoollist{#2}%
  \end{minipage}}
\newlength{\iatoolwidth}

\begin{table*}[h]
\centering
\caption{\textbf{InterA11y tools and corresponding success criteria.}}
\label{tab:intera11y-tools}
\small
\setlength{\tabcolsep}{3pt}
\setlength{\iatoolwidth}{\dimexpr\textwidth-6\tabcolsep\relax}
\renewcommand{\arraystretch}{1.08}
\begin{tabular}{@{}
  p{.10\iatoolwidth}
  p{.53\iatoolwidth}
  p{.27\iatoolwidth}
  p{.1\iatoolwidth}@{}}
\hline
\textbf{Category} & \textbf{Scan tools} & \textbf{Agentic tools} & \textbf{SC} \\
\hline
    \rowcolor{gray!8}
Inspection &
\iatwoscans{\iatool{inventory} (links)\\
  \iatool{inventory} (images/media)}{\iatool{inventory} (lookalike glyphs)\\
  \iatool{run} (at-rest errors)} &
\iatoollist{\iatool{query\_\allowbreak ax\_\allowbreak node}\\
  \iatool{resolve\_\allowbreak destination}} &
\iatoollist{1.1.1, 1.4.1\\
  2.4.4, 3.3.1\\
  4.1.3} \\
Capture &
\iatwoscans{\iatool{captureIndicator}\\
  \iatool{sequenceMap}}{\iatool{cropOf} (graphics)} &
\iatoollist{\iatool{request\_\allowbreak hi\_\allowbreak res\_\allowbreak crop}\\
  \iatool{set\_\allowbreak state\_\allowbreak and\_\allowbreak capture}\\
  \iatool{capture\_\allowbreak full\_\allowbreak page}} &
\iatoollist{1.1.1, 1.4.1\\
  2.4.3, 2.4.7} \\
    \rowcolor{gray!8}
Measurement &
\iatwoscans{\iatool{colours} (text contrast)\\
  \iatool{linkCues}\\
  \iatool{styleDiff} (state peers)}{\iatool{styleDiff} (form labels)\\
  \iatool{styleDiff} (colour-coded siblings)\\
  \iatool{overlapsOtherContent}} &
\iatoollist{\iatool{measure\_\allowbreak geometry\_\allowbreak live}\\
  \iatool{compute\_\allowbreak contrast\_\allowbreak ratio}\\
  \iatool{resolve\_\allowbreak part\_\allowbreak color}\\
  \iatool{compare\_\allowbreak named\_\allowbreak regions}} &
\iatoollist{1.1.1, 1.4.1\\
  1.4.13, 2.4.7} \\
Interaction &
\iatwoscans{\iatool{walk} (Tab)\\
  \iatool{walk} (Shift+Tab)\\
  \iatool{detectKeyboardTraps}\\
  \iatool{detectFocusRetentionTraps}\\
  \iatool{detectFixedSetConfinementTraps}\\
  \iatool{detectFocusRejection}\\
  \iatool{compositeWalk}\\
  \iatool{activateOnce} (keyboard)\\
  \iatool{activateOnce} (pointer)}{\iatool{keyboardInRevealed}\\
  \iatool{typeInto}\\
  \iatool{hoverReveal}\\
  \iatool{testTrigger} (pointer travel)\\
  \iatool{testTrigger} (Escape)\\
  \iatool{testTrigger} (focus parity)\\
  \iatool{scenario} (empty input)\\
  \iatool{scenario} (invalid input)\\
  \iatool{scenario} (valid input)} &
\iatoollist{\iatool{interact\_\allowbreak and\_\allowbreak observe}\\
  \iatool{render\_\allowbreak with\_\allowbreak overrides}} &
\iatoollist{1.4.1, 1.4.13\\
  2.1.1, 2.1.2\\
  2.4.3, 2.4.4\\
  2.4.7, 3.3.1\\
  4.1.3} \\
    \rowcolor{gray!8}
Monitoring &
\iatwoscans{\iatool{idleWatch}}{\iatool{testTrigger} (persistence)} &
\iatoollist{\iatool{observe\_\allowbreak state\_\allowbreak after\_\allowbreak activation}\\
  \iatool{probe\_\allowbreak screen\_\allowbreak reader\_\allowbreak after\_\allowbreak action}} &
\iatoollist{1.4.13, 2.1.1\\
  2.1.2, 2.4.3\\
  3.3.1, 4.1.3} \\
\hline
\textbf{Total} & \textbf{33} & \textbf{13} & \\
\hline
\end{tabular}
\end{table*}

\begin{table*}[h]
  \centering
  \small
  \caption{Expert evaluated websites' URLs. Parentheses show the number of pages included from the domain if greater than 1.}
  \label{tab:website-areas}
  \begin{tabularx}{0.9\textwidth}{
    @{}
    >{\raggedright\arraybackslash}p{0.27\textwidth}
    >{\raggedright\arraybackslash}X
    @{}
  }
    \toprule
    \textbf{Area} & \textbf{Base URLs} \\
    \midrule
    \rowcolor{gray!8}
    General Retail and Department Stores
      & amazon.com (3), temu.com \\
    Fashion and Apparel
      & boohooman.com, gymshark.com, www2.hm.com, macys.com,
        nordstrom.com, reebok.com \\
    \rowcolor{gray!8}
    Home and Lifestyle Retail
      & wayfair.com \\
    Restaurants
      & dominos.com, panerabread.com, sweetgreen.com \\
    \rowcolor{gray!8}
    Travel and Transport
      & iflysouthern.com, hotels.spirit.com, vueling.com \\
    Technology and Electronics Retail
      & newegg.com \\
    \rowcolor{gray!8}
    Media and Entertainment
      & bbc.com, buzzfeed.com, espn.com, nymag.com,
        rottentomatoes.com \\
    Education and Learning
      & doomersion.com, kahoot.com, quizlet.com \\
    \rowcolor{gray!8}
    Finance and Investment
      & etfdb.com, wallstreetoasis.com, finance.yahoo.com (2) \\
    Real Estate
      & opendoor.com, zillow.com \\
    \rowcolor{gray!8}
    Social Networks and Communities
      & linkedin.com, quora.com, reddit.com (3) \\
    Government and Public Services
      & vitalchek.com \\
    \rowcolor{gray!8}
    Industrial and Agricultural Services
      & artera.com, grazemate.com, morsecorp.com \\
    Business Software and Services
      & calendly.com, cloudflare.com, fed10.ai, drive.google.com,
        jobs.ashbyhq.com (2), ix.dev, klaviyo.com, notion.com,
        o11.ai, openroll.com, reacherapp.com, therubric.ai \\
    \bottomrule
  \end{tabularx}
\end{table*}

